\documentstyle[12pt]{article}
\newcommand{\bce}{\begin{center}}
\newcommand{\ece}{\end{center}}
\newcommand{\beq}{\begin{equation}}
\newcommand{\eeq}{\end{equation}}
\newcommand{\bea}{\vspace{0.25cm}\begin{eqnarray}}
\newcommand{\eea}{\end{eqnarray}}

\newcommand{\bsigma}{\mbox{\boldmath $\sigma$}}

\newcommand{\ba}{\begin{array}}
\newcommand{\ea}{\end{array}}

\def\lsim{\mathrel{\rlap{\lower4pt\hbox{\hskip1pt$\sim$}}
    \raise1pt\hbox{$<$}}}	  
\def\gsim{\mathrel{\rlap{\lower4pt\hbox{\hskip1pt$\sim$}}
    \raise1pt\hbox{$>$}}}	  

\def\lsim{\mathrel{\rlap{\lower4pt\hbox{\hskip1pt$\sim$}}
    \raise1pt\hbox{$<$}}}         
\def\gsim{\mathrel{\rlap{\lower4pt\hbox{\hskip1pt$\sim$}}
    \raise1pt\hbox{$>$}}}         

  \advance\oddsidemargin  by -1in
  \advance\evensidemargin by -1in
  \advance\topmargin      by -0.5in
\def\lsim{\mathrel{\rlap{\lower4pt\hbox{\hskip1pt$\sim$}}
    \raise1pt\hbox{$<$}}}         
\def\gsim{\mathrel{\rlap{\lower4pt\hbox{\hskip1pt$\sim$}}
    \raise1pt\hbox{$>$}}}         

\def\beq{\begin{equation}}
\def\endeq{\end{equation}}
\def\arr{\begin{eqnarray}}
\def\endarr{\end{eqnarray}}
\makeindex
 
\begin{document}

\phantom.\hspace{8.8cm}{\large\bf KFA-IKP(TH)-1996-16\bigskip\\}
\phantom.\hspace{10.2cm}{\large \bf 29 October   1996}\vspace{1.5cm}\\
\begin{center}
{\Large \bf
Rare processes and coherent phenomena in crystals
 \vspace{1.0cm}}
 
{\large \bf
V.R.Zoller\medskip\\ }
{\large \it
Institut  f\"ur Kernphysik, Forschungszentrum J\"ulich,\\
D-52425 J\"ulich, Germany {\footnote {KPH166@AIX.SP.KFA-JUELICH.DE}}\medskip\\
Institute for Theoretical and Experimental Physics,\\
ul. B.Chermushkinskaya 25, 117218 Moscow, Russia
{\footnote { ZOLLER@HERON.ITEP.RU}}\vspace{1cm}\\}
{\bf           Abstract}
\end{center}
We study coherent enhancement of  Coulomb excitation of 
 high energy  particles in crystals.
We develop multiple scattering theory description of 
 coherent excitation  which consistently incorporates  both the 
specific resonant properties of particle-crystal interactions and the final/initial
 state interaction  effects 
typical of the diffractive scattering.
Possible applications  
to observation of  induced radiative 
neutrino transitions are discussed.

\newpage

\section{} 

In a two-level quantum    system (the ground state $ |0\rangle$
 and the excited state $|1\rangle$)   under periodical perturbation
$V\sin{\nu t}$ with the frequency $\nu$ equal to
the level splitting  $\nu_{10}=E_1-E_0$ there develop quantum beats with
the oscillation frequency $\omega= \langle 1|V|0\rangle$. 
 If the perturbation $V$ is weak, then for $\omega t\ll 1$
the $|0\rangle\to |1\rangle$ the transition probability $P_{10}(t)$ 
increases rapidly with the time
\beq
P_{10}(t)\,\propto\,\omega^2t^2\,.        \label{eq:1}
\endeq  
 
In particle physics, examples of such rare processes are
 the weak radiative transitions of hyperons, $N\gamma\to Y$, 
 beyond-the-standard-model-decays like $\mu\to e\gamma$ and radiative
(magnetic) neutrino conversion $\nu_1\gamma\to \nu_2$. 
For instance, if one could 
subject hyperons to  a high frequency field, 
$\nu\sim m_Y-m_N$,
 the rates of rare  decays can be enhanced substantially.
The high  monochromaticity
is an evident condition to sustain the growth (\ref{eq:1}) over large 
time scale.

Okorokov \cite{O1},\cite{O2},\cite{O3}
 was the first to suggest that all of the  above requirements are met
 best 
in Coulomb interaction of a  
high-energy particle propagating in a crystal along
 the crystallographic axis. Here the r\^ole of "time" is  played by
 the crystal thickness $ L$.
 In \cite{O1} the  resonant excitation of atoms and
 nuclei by a periodical Coulomb field of a crystal  was predicted.
 In \cite{O2} the first
 observation of the resonant transition of $He^+(n=1)$ to  $He^+(n=4)$ in $Ag$ crystal
 was reported. For a propagation through a crystal with the lattice spacing
$d$, the frequency $\nu=2\pi{v/d}$, 
where $v$ is the velocity of the atom. For ultrarelativistic particles $\nu$ is 
enhanced due to
 the Doppler shift, $\nu=2\pi\,\gamma\,{v/d}$, where $\gamma$ is the Lorentz
factor. It is the Lorentz factor which can boost $\nu$ to the hundreds $MeV's$ range.
 Since \cite{O1} the   Okorokov effect
 has been studied extensively both experimentally
\cite{EXP} and theoretically \cite{TH}.

In early works on the subject
the Coulomb field of a crystal was evaluated in
  the Weizs\"aker-Williams approximation  \cite{T-M}, \cite{FER} 
and then applied to the calculation of 
the transition amplitude in the plane wave  Born approximation.
 The $N\gamma\to Y$ transitions in the Coulomb field of the nucleus were  
discussed by Pomeranchuck and
 Shmushkevich\cite{P-SH} .
 The $p\gamma\to \Sigma^+$ excitation in crystals 
based on the approach  \cite{T-M}, \cite{FER} and \cite{P-SH} 
was considered recently in \cite{DUB} and 
it was claimed that the law  
$
P_{\Sigma p}(N)\,\propto\,N^2\,,
$ 
  holds  up to the
 crystal thicknesses  $N=L/d\simeq 10^7$ \cite{DUB}. 

However, the plane wave Born approximation is not self-contained, it does not allow
to assess what is the upper limit on $N$ and, as a matter of fact, it 
grossly overestimates the enhancement factor. Our point is that the coherency
of $p\gamma\to\Sigma^+$ transitions depends on the initial state interactions (ISI)
 of
 the proton and final state interactions (FSI) of the hyperon. It also depends on 
thermal vibrations of the atoms in a crystal.

The purpose  of this communication  is to derive the Okorokov effect for the 
Pomeranchuck-Shmushkevich
processes directly from the multiple scattering (MS) theory.
We find a dramatic impact of 
the ISI and FSI effects 
on
$p\to\Sigma^+$-transition   resulting in  the
very slow, $\sim \log N$, rise of the  amplitude
rather than $\sim N$.
 This result is quite general and is equally applicable to
the excitation of ultrarelativistic nuclei and ions passing through the crystals.

 On the contrary, the 
 magnetic conversion of the neutrino in crystals as shown
to have  a pronounced resonant structure and we comment on possible implications 
 for the future laboratory investigations 
of the neutrino electrodynamics.

\section{}

Consider Coulomb interaction of high-energy particle $a$  moving along the 
  $<001>$ axis in
 monatomic crystal.  
The interatomic distances   are large, $d\sim 5 \,\AA$, 
compared to the  
 Thomas-Fermi screening radius $a_{TF}$,
 $a_{TF}=\mu^{-1}=(m_e \alpha Z^{1/3})^{-1}\simeq 0.5Z^{-1/3}\,\AA$,
 where $Z$ is the atomic number and $\alpha=1/137$.
The relevant impact parameters, $b_{eff}$, 
satisfy $ b_{eff}\sim \mu^{-1}\ll d$
and the widely used  
one-chain approximation is applicable. 
At the same time, $b_{eff}$ is much larger 
than the nuclear radius
$R_A\sim 10^{-5}Z^{1/3}\AA$ and the effects of strong interactions
 can be safely neglected. 
  
The amplitude ${\cal F}_{ba}$ of coherent transition $a\to b$ on a chain 
 of
 $N$ identical atoms is 
\beq
{\cal F}_{ba}=
\langle\Psi({\bf r}_1,...,{\bf r}_N)|S_{ba}|\Psi({\bf r}_1,...,{\bf r}_N)\rangle\,.
							\label{eq:T}
\endeq
Here $\Psi({\bf r}_1,...,{\bf r}_N)$ is the full ground-state wave function of the 
 $N$-atomic chain and ${\bf r}_{i}=({\bf b}_i,z_i)$ are positions of atoms.
 The S-matrix,
 $S_{ba}$, admits the expansion into  multiple scattering series,
${\cal F}_{ba}=\sum_n F^{(n)}_{ba}$, and
 describes the $a$-to-$b$ transition as well as
 both the initial and final state interactions
 of both particles $a$ and  $b$ with crystal. 
The high momentum projectile propagates along 
 straight-line trajectories at fixed
 impact parameter
$\bf b$. 
Then one can use Gribov's dispersion integral
representation for the coupled channel
$n$-fold scattering amplitude $ F^{(n)}_{ba}({\bf b})$  in the basis of physical states
 $|a\rangle,|i\rangle,...,|b\rangle$ \cite{GRIB}. This allows one to consider the  
$a\to b$ transition as a sequence of  $n$ both diagonal and off-diagonal transitions 
$|a\rangle\to|i\rangle\to...|k\rangle\to|j\rangle\to|b\rangle$ ordered along
the beam direction.
 Each off-diagonal transition, 
involves a longitudinal momentum transfer \cite{GRIB}
\beq
\kappa_{ji}={{\Delta m^2_{ji}}\over {2E}}\,, \label{eq:KAP}
\endeq
where  $\Delta m^2_{ji}=m^2_j-m^2_i$,
 $E$ is the projectile energy and $m_j$ is the particle $j$ mass. Associated with
the $\kappa_{ji}$ is the phase factor
$\exp[i\kappa_{bj}z_{\nu}+i\kappa_{jk}z_{\nu-1}+...+i\kappa_{ia}z_{1}],$
 in   the amplitude $F^{(n)}_{ba}({\bf b})$:
\arr
F^{(n)}_{ba}({\bf b})=(-i)^{n-1}{N!\over (N-n)!}\sum_{i,...,k,j}
\int_{0}^{L}dz_n\int d^2{\bf b}_{n}f_{bj}({\bf b}-{\bf b}_{n})
\exp[i\kappa_{bj}z_n]                                                            \\ \nonumber
\int_{0}^{z_n}dz_{n-1}\int d^2{\bf b}_{{n-1}}f_{jk}({\bf b}-{\bf b}_{n-1})
\exp[i\kappa_{jk}z_{n-1}]...                                                      \\ \nonumber
...\int_{0}^{z_2}dz_{1}\int d^2{\bf b}_{{1}}f_{ia}({\bf b}-{\bf b}_{1})
\exp[i\kappa_{ia}z_{1}]\chi({\bf r}_1,...,{\bf r}_n) \, , \label{eq:Fn}
\endarr
where 
\beq
\chi({\bf r}_1,...,{\bf r}_n)=\int d^3{\bf r}_{n+1}...d^3{\bf r}_{N}
|\Psi({\bf r}_1,...,{\bf r}_N)|^2 \, .                     \label{eq:CHI}
\endeq
For our purposes it is sufficient to use 
 the uncorrelated wave function
\beq
|\Psi({\bf r}_1,...,{\bf r}_N)|^2=\prod_{i=1}^N \phi({\bf u}_i)\,, \label{eq:Psi}
\endeq
where
 the normal coordinates ${\bf u}_i$  are defined by 
 ${\bf r}_i=(i-1){\bf d}+{\bf u}_i$, $i=1,...,N$,
  ${\bf d}=(0,\,0,\,d)$ and ${\bf u}_i=({\bf b}_i,u_{zi})$.

For the rare processes of interest 
\beq
f_{ij}\ll f_{ii},f_{jj}\,\,,i\neq j
\endeq
and  we keep only the lowest order 
terms in powers of  the off-diagonal transitions.
Then,
\beq
{\cal F}_{ba}({\bf b})=\sum^N_{n=1}F^{(n)}_{ba}({\bf b}) 
=\sum_{m=1}^{N}\sum_{n_1=0}^{m-1}\sum_{n_2=0}^{N-m}
h_{aa}({\bf b})^{n_1}
h_{bb}({\bf b})^{n_2}
h_{ba}({\bf b})S^{(m)}_L(\kappa_{ba})\, ,\label{eq:Fsum}
\endeq
where
\beq
h_{ba}({\bf b})=\int d^2{\bf b}_{i}
f_{ba}({\bf b}-{\bf b}_{i})\rho({\bf b}_{i})\, , \label{eq:HBA}
\endeq
\beq
\rho({\bf b}_{i})=\int du_{zi} \phi({\bf u}_i)={3\over {2\pi\langle{\bf u}^2\rangle}}
\exp\left[-{3{\bf b}^2_{i}\over
 {2\langle{\bf u}^2\rangle}}\right]\, , \label{eq:RHO}
\endeq
\beq
S^{(m)}_L(\kappa_{ba})=\int^{L}_{0} du_{zm} \phi({\bf u}_m)\exp(i\kappa_{ba}u_{zm})
=\exp\left[-{1\over 6}\kappa^2_{ba}\langle{\bf u}^2\rangle\right]
\exp\left[i\kappa_{ba}(m-1)d\right]\, .
\endeq
 For the sake of simplicity, we have assumed the Gaussian form of
$\phi({\bf u}_i)$. In (\ref{eq:RHO})
 $\langle{\bf u}^2\rangle$ is
the mean squared amplitude
of thermal vibrations of the lattice .
 Note 
 that 
 the relevant longitudinal momenta 
   are as follows  $\kappa_{ji}\sim d^{-1}\ll\langle{\bf u}^2\rangle^{-1/2} $.
The summation over $m$ in eq.(\ref{eq:Fsum})
 results in
\arr
{\cal F}_{ba}({\bf b})=
 h_{ba}({\bf b})
\exp\left[-{1\over 6}\kappa^2_{ba}\langle{\bf u}^2\rangle\right]\\ \nonumber
\times{{[1+ih_{bb}({\bf b})]^N-
[1+ih_{aa}({\bf b})]^N\exp\left[i\kappa_{ba}Nd\right]}\over
{[1+ih_{bb}({\bf b})]-
[1+ih_{aa}({\bf b})]\exp\left[i\kappa_{ba}d\right]}}\,. \label{eq:FGEN}
\endarr
In the  practically interesting case like $p\gamma\to\Sigma^+$ and/or $e\gamma\to\mu$
we have $f_{aa}=f_{bb}$ and the amplitube ${\cal F}_{ba}$ reads
\beq
{\cal F}_{ba}({\bf b})=
 h_{ba}({\bf b})
\left[1+ih_{aa}({\bf b})\right]^{N-1}S_L(\kappa_{ba})
  \,,                                                   \label{eq:FSIM}
\endeq   
where $S_L(\kappa)$ is the longitudinal form factor 
\beq
S_L(\kappa_{ba})=\exp\left[-{1\over 6}\kappa^2_{ba}\langle{\bf u}^2\rangle\right]
{{\sin(\kappa_{ba}Nd/2)}\over{\sin(\kappa_{ba}d/2)}}\,.
					\label{eq:SL}
\endeq
Evidently, for  the resonant
values of $\kappa_{ba}$, 
\beq
 \kappa_{ba}={2\pi n\over d}\,,\,n=0,\pm 1,\pm 2...\,,\label{eq:KAPRES}
\endeq
we have $S_L\propto N$. The resonance energy
 $E_n$, 
\beq
E_n={\Delta m^2_{ba}d\over 4\pi n}   \label{eq:ERES}
\endeq
can be lowered going to $n\gg1$. However, there is an upper limit on $n$
\beq
n_{max}\simeq {1\over 2\pi}\left({3d^2\over \langle{\bf u}^2\rangle }\right)^{1/2}
\simeq 5-10\,.                              \label{eq:NMAX}
\endeq
which comes from the Debye-Waller factor
 in $S_L(\kappa)$, eq.(\ref{eq:SL}).

Above we focused on the S-matrix at a fixed impact parameter ${\bf b}$.
The experimentally observable quantity is the differential cross section
$d\sigma_{ba}/dq^2_{\perp}=|{\cal T}_{ba}({\bf q}_{\perp})|^2$, where
\beq
{\cal T}_{ba}({\bf q_{\perp}})=S_L(\kappa_{ba})
\int {d^2{\bf b}\over {2\pi}}\,
 h_{ba}({\bf b})
\left[1+ih_{aa}({\bf b})\right]^{N-1}
\exp[i{\bf q_{\perp}}{\bf b}]\,\label{eq:TDEF}
\endeq
and  ${\bf q_{\perp}}$ - is the two-dimensional vector of the transverse momentum 
($|{\bf q_{\perp}}|=q_{\perp}$).

As we shall see below, the dominant contribution to  (\ref{eq:TDEF}) comes
from $b_{eff}\sim \mu^{-1}N_{eff}$, where 
$$
N_{eff}=\log(Z\alpha N)\gg 1 \,.
$$
Then, because $\langle{\bf u}^2\rangle\sim \mu^{-2}$
\beq
h_{aa}({\bf b})=\int d^2{\bf b}_{i}
f_{aa}({\bf b}-{\bf b}_{i})\rho({\bf b}_{i})
  \simeq f_{aa}(\bf b)\,.  \label{eq:HBAFAA}
\endeq 
The amplitude $f_{aa}$ of elastic scattering of
 a charged  particle $a$ in a screened Coulomb
 field of an
 isolated atom is 
purely real if $q^2_{\perp}\ll \mu p$,
 where $p$ is the projectile momentum:
\beq
f_{aa}(b)  =Z\alpha K_0(\mu b)\,. \label{eq:FCOUL} 
\endeq
Here we neglect the contributions
of  anomalous 
 magnetic moments to the small-angle elastic scattering.

Let the resonance condition (\ref{eq:KAPRES}) be satisfyed
 so that $S_L(\kappa)\propto N$.
Our point is that because of the distortion factor $[1+ih_{aa}({\bf b})]^{N}$
in the integrand of eq.(\ref{eq:TDEF}) the law ${\cal T}_{ba}\propto N$ does not hold.
 The evaluation of the distortion effects for the $p\gamma\to\Sigma^+$ transition 
proceeds as follows.   

For 
 the phenomenological transition  matrix element \cite{P-SH}, \cite{K-SH}
\beq
{\cal M}(p\gamma\to\Sigma^+)=i\bar u_{\Sigma} 
\left(\mu_{\Sigma p}+\gamma_5 d_{\Sigma p}\right)
\sigma_{\mu\nu} q^{\nu}\varepsilon^{\mu}u_p \label{eq:M}
\endeq
one readily finds
\beq
{f}_{\Sigma p}({\bf b})
=\left[\mu_{\Sigma p}\bsigma\left[{\bf b}\times{\bf n}\right] 
+id_{\Sigma p}\bsigma{\bf b}\right]
b^{-1}Z\sqrt{\alpha}\mu K_1(\mu b)\,.                   \label{eq:FSIGMA}
\endeq  
Here
${\bf n}$ -
 is a unit vector along the projectile direction,
${\bsigma}$ is the Pauli spin vector and  $K_1(x)$ is a
 modified Bessel function.

Then, the evaluation of the  full transition amplitude reads
\arr
{\cal T}_{\Sigma p}({\bf q_{\perp}})=
\left[ i\mu_{\Sigma p}{\bsigma}\left[{\bf q_{\perp}}\times{\bf n}\right]+
d_{\Sigma p}{\bsigma}{\bf q_{\perp}}\right]q_{\perp}^{-1}
\mu S_L(\kappa_{ba})                                             \\   \nonumber
\times Z\sqrt{\alpha}\int bdb\,J_1(q_{\perp}b)K_1(\mu b))\left[1+iZ\alpha K_0(\mu b)\right]^{N-1}
\label{eq:TSIGMA} 
\endarr 
and the  steepest descent from the  saddle-point  at
$\mu b=N_{eff}-i\pi/2$
 yields for  $q_{\perp}\lsim\mu N_{eff}^{-1}$
\beq
{\cal T}_{\Sigma p}({ q}_{\perp})\propto
i\, (N_{eff}-i\pi/2)\, J_1\left(q_{\perp}\mu^{-1}(N_{eff}-i\pi/2)\right)\,.
                                                              \label{eq:TESTIM}
\endeq
where 
  $J_1(x)$ is the  Bessel function. Consequently, as soon as $N\alpha Z\gg1$, which holds
for all practical purposes, even under the resonance condition (\ref{eq:KAPRES}) one
only has the logarithmic growth
\beq
{\cal T}_{\Sigma p}\propto \log(Z\alpha N)\,.
\endeq
Although the relevant impact parameters rise, $b_{eff}\sim\mu^{-1}N_{eff}$,
 for all the practical
purposes $b_{eff}\ll d$ so that the one-chain approximation holds.
The total cross-section,
$
\sigma=\int {d{ q_{\perp}^2}}|{\cal T}_{\Sigma p}(q_{\perp})|^2
$
 appeares to be  independent of $N$.

We conclude that  ISI and FSI effects completely destroy the
 resonant enhancement effect. Similar phenomena are  well known in the high-energy 
diffractive scattering of hadrons, though for purely imaginary  amplitudes.
For the  screening
of real elastic amplitudes in the Coulomb potential scattering see e.g.   
  \cite{K-M}.

\section{} 

Several proposals \cite{GON} of  experiments
 on 
the neutrino conversion induced by the neutrino magnetic moment  in 
external electromagnetic  field
\beq
\nu_1\gamma\to\nu_2 \label{eq:NUTONU}
\endeq 
are under discussion \cite{FRERE}. Hereafter we are dealing with
 the non-diagonal magnetic ($\mu_{21}$) and electric ($d_{21}$) dipole 
 moments which exist 
 in general for both Dirac and Majorana neutrinos \cite{VVO}.

The experimental bounds on $\mu_{21}$ are such \cite{FRERE} that
the FSI/ISI effects are negligible for all 
practical purposes.
Then, in terms of the  matrix element (\ref{eq:M}) 
the differential cross section of the process (\ref{eq:NUTONU}) reads
\beq 
{d\sigma_{21}\over dq_{\perp}^2}=\alpha Z^2
b^2_{21}
{q_{\perp}^2\over ({q_{\perp}^2+\mu^2})^2}S^2_T(q_{\perp}^2)S^2_L(\kappa_{21})\,,
                                                                       \label{eq:DIF}
\endeq 
where
$b^2_{21}=|\mu_{21}|^2+|d_{21}|^2$ for Dirac neutrinos and 
$b^2_{21}=4\left[({\rm Im}\mu_{21})^2+({\rm Re} d_{21})^2\right]$
for Majorana neutrinos.
The transverse form factor $S_T(q_{\perp}^2)$ equals 
\beq
S_T(q_{\perp}^2)=\exp\left[-{1\over 3}q^2_{\perp}\langle{\bf u}^2\rangle\right].
\endeq

The observation of the $\nu_1\gamma\to\nu_2$-transition which would look
 like a
 resonance 
at some neutrino energy $E\simeq E_{n}$ (see  eq.(\ref{eq:ERES}))
would  enable one to evaluate 
the mass difference of the  neutrinos of different species
and estimate the neutrino transition magnetic moment.

 For these purposes it is useful
to represent the resonant part 
of the differential cross section (\ref{eq:DIF}) 
as a sum of Lorentz curves. Then, the neutrino conversion rate,
 $R_c=\sigma_{21}/d^2$, equals
\beq
R_c\simeq N^2
{b^2_{21}\alpha Z^2\over {d^2\left[1+{2\over 3}\mu^2\langle{\bf u}^2\rangle\right]^2}}
 \sum_n {(\Gamma_n/2)^2\over {(\Gamma_n/2)^2}+(E-E_n)^2}
\exp\left[-{4\pi^2n^2\over 3}{\langle{\bf u}^2\rangle\over d^2}\right]
\,.
\label{eq:LOR}
\endeq 
 The  width of $n$-th resonance is 
\beq
\Gamma_n={\sqrt{3}\over 2}
{\Delta m^2_{21}d\over {\pi^2 n^2 N}}
\endeq

Then, the "disappearance type" 
experiment \cite{GON}  with
 the CERN  $30\,GeV$ $\nu$-beam supplemented with the crystal target 
would allow to investigate the region of $\Delta m^2_{21}$ up to $\sim (1\,MeV)^2$.
If we accept that the departure of the neutrino from the crystallographic axis which
does destroy the  coherence is due to the transverse momenta in the range 
from $q_{\perp}\sim d^{-1}$ up to $q_{\perp}\sim \mu$, the crystal-target experiment
could hope to detect transition 
neutrino magnetic moments larger than
\beq
\mu_{21}\sim (10^{-6}-10^{-8})\mu_B, \label{eq:MUB}
\endeq
where $\mu_B$ is Bohr magneton. The efficiency is assumed to be $\sim 10^{-4}$
 \cite{GON}. 
 It should be
 emphasized that in the  
$\nu-crystal$ coherent interaction
 typical $q^2$
vary in the range  $\sim (10\,KeV)^2$. For comparison, 
presently available bounds on $\mu_{21}$ were obtained 
for virtual photons with typical
$q^2\sim (100\,MeV)^2$.
 A nontrivial $q^2$
 dependence,
resulting in a strong low-momentum enhancement of
 a neutrino magnetic form factor, discussed in \cite{FRERE}
makes such an experiment topical.\\

{\bf Acknowledgements:} The author is grateful to
J.Speth for the hospitality at the Institut f\"ur Kernphysik, KFA
J\"ulich.  Thanks are due to N.N. Nikolaev for
careful reading the manuscript and helpful comments.
 Useful discussions with V.V. Okorokov and correspondence with  M.I. Vysotsky
 are greatfully acknowledged. 

\newpage\


\begin{thebibliography}{99}
 
\bibitem{O1}
V.V.Okorokov, Sov. J. Nucl. Phys. 2(1965) 719; JETP Lett. 2 (1965) 111.
\bibitem{O2}
V.V. Okorokov, D.L Tolchenkov, Yu.P. Cheblukov et al., JETP Lett 16 (1972); Phys.
Lett. A43 (1973) 485.
\bibitem{O3}
V.V. Okorokov, JETP Lett., 62 (1995) 911. 
\bibitem{EXP}
M.J.Gaillard, J.C.Poizat, J.Remillieux, and M.L.Gaillard, Phys. Lett.
 A45 (1973) 306;\\
H.G. Berry, D.S. Gemmel, R.E. Holland et al., Phys. Lett. A49 (1975) 123;\\
M. Mannami, H. Kudo, M. Matsushita and K. Ishii, Phys. Lett. A64 (1977) 136;\\
S. Datz, C.M. Moak, O.H. Crawford et al., Phys. Rev. Lett. 40 (1987) 843;\\
C.M. Moak, S. Datz, O.H. Crawford et al., Phys. Rev. A19 (1979) 977;\\
F. Fujimoto, Nucl. Instr. Methods B40/41 (1989) 165;\\
Y. Iwata, K. Komaki, Y. Yamazaki et al., Nucl. Instr. Methods, 48 (1990) 163.
\bibitem{TH} 
J. Kondo, J.Phys. Soc. Jpn., 36 (1974) 1406;\\
S. Shindo and Y.H. Ohtsuki, Phys. Rev. B14 (1976) 3929;\\
Y. Yamashita and Y.H. Ohtsuki, Phys. Rev. B22 (1980) 1183;\\
Yu.L. Pivovarov and A.A. Shirokov, Sov. J. Nucl. Phys. 37 (1983) 653.
\bibitem{T-M} 
M.L. Ter-Mikaelyan, Sov. Phys. JETP, 25 (1953) 289.
\bibitem{FER}
 B. Ferretti, Nuovo Cimento, 7 (1950) 1; B7 (1972) 225; B9 (1972) 399.
\bibitem{P-SH}
I.Ya. Pomeranchuck and I.M. Shmushkevich, Nucl. Phys. 23 (1961) 452.
\bibitem{DUB}
A.Yu. Dubin, Sov.J.Nucl.Phys. 52 (1990) 790.
\bibitem{GRIB}
V.N. Gribov, Sov. Phys. JETP 29 (1969) 483; 30 (1970) 709.
\bibitem{K-SH}
R.F. Behrends, Phys. Rev. 111 (1958) 1691;\\
V.I. Zakharov and A.B. Kaidalov, Sov. J. Nucl. Phys. 5 (1967) 259.
\bibitem{K-M}
L.I. Schiff, Phys. Rev. 103 (1956) 443;\\
N.P. Kalashnikov and V.D. Mur, Sov. J. Nucl. Phys. 16 (1973) 613.
\bibitem{GON}
M.C. Gonzalez-Garcia, F. Vannucci and J. Castromonte, Phys. Lett. B373 (1996) 153;\\
S. Matsuki and K. Yamamoto, Phys. Lett. B289 (1992) 194;\\
Makoto Sukuda, Phys. Rev. Lett. 72 (1994) 804;\\
J. Bernabeu, S.M. Bilenkii, F.G. Botella and J. Segura, Nucl. Phys. B426 (1994) 434.
\bibitem{FRERE}
J.-M. Frere, R.B. Nevzorov and M.I. Vysotsky, Preprint ULB-TH/96/14. 
\bibitem{VVO}
J. Schechter and J.W.F. Valle, Phys. Rev. D24 (1981);\\
P.B. Pal and L. Wolfenstein, Phys. Rev. D25 (1982);\\
J. Nieves, Phys. Rev. D26 (1982);\\
M.B.Voloshin and M.I.Vysotsky,  Sov. J. Nucl. Phys. 44(1986) 544;\\
M.B.Voloshin, M.I.Vysotsky and L.B. Okun', Sov. Phys. JETP 64 (1986) 446.  

\end{thebibliography}
\end{document}